\documentclass{IEEEtran}
\usepackage{cite}
\usepackage{amsmath,amssymb,amsfonts}
\usepackage{algorithmic}
\usepackage{graphicx}
\usepackage{textcomp}
\usepackage{mathtools}
\usepackage{xcolor}
\def\BibTeX{{\rm B\kern-.05em{\sc i\kern-.025em b}\kern-.08em
    T\kern-.1667em\lower.7ex\hbox{E}\kern-.125emX}}
\begin{document}
\title{A Note on Surgical Eigenstructure Assignment via State Feedback}
\author{Abdullah Al Maruf and Sandip Roy %\IEEEmembership{Member, IEEE}
\thanks{This work was partially supported by NSF Grant 1635184.}
\thanks{Authors are with School of Electrical Engineering and Computer Science, Washington State University, Pullman, WA 99164, USA. (email:abdullahal.maruf@wsu.edu, sandip@wsu.edu)}}

\maketitle

\begin{abstract}
A method for assigning all eigenvalues and a subset of key eigenvectors/generalized eigenvectors of a linear time-invariant system via state feedback is developed, which is complete in the sense that it encompasses the repeated-eigenvalue case. Our analysis shows that it is sufficient for the key eigenvectors/generalized eigenvectors and their associated eigenvalues to lie within the subspaces identified in the classical eigenstructure assignment literature, whereupon the remaining eigenvalues can be assigned at will. The developed method is computationally attractive and also provides a constructive proof for the assignability condition. Finally, a generalization of the problem where some eigenvalues may also be left unspecified is addressed. 
\end{abstract}


\section{Introduction}

\label{sec:introduction}

Eigenstructure assignment, which is concerned with placement of both eigenvalues and eigenvectors of a linear time-invariant system through feedback, has been comprehensively addressed in the controls-engineering literature over the last 50 years \cite{b1,b2,b3,b4,b5,b6,g1,g2}. The important early work of Moore identifies conditions on the full spectrum for assignability using state feedback, and also provides an algorithm for construction of the feedback control \cite{b2}.  Subsequent studies have: 1) generalized this analysis to encompass the defective-eigenvalue case, 2) obtained alternative characterizations of assignability, 3) developed algorithms for assignment that are computationally appealing, 4) addressed the output-feedback control problem, and 5) used the geometric control theory to further characterize the assignable eigenvector directions, among other directions.

Many large-scale systems do not require full eigenstructure assignment, nor do they have sufficiently rich actuation to allow for useful flexibility in shaping the full eigenstructure. In some of these settings, control designers instead seek to assign all eigenvalues but only a subset of key eigenvectors of the system.  Problems of this sort arise, for instance, in malicious control of the power grid and secure management of multi-agent systems \cite{jointeig_app1,jointeig_app2,jointeig_app3}. The problem of assigning all eigenvalues and specified subset of eigenvectors of a linear time-invariant system via state feedback -- which we refer to as the {\em surgical eigenstructure assignment} problem -- should admit a solution under more relaxed conditions than the full eigenstructure problem, and also allow for computationally simpler algorithms. A general solution does not, however, follow immediately from the methods for full eigenstructure assignment \cite{b2,b3},  because of inclarity about whether and how the unspecified eigenvectors  can be chosen to meet the criteria for assignability. 

A separate track of research on partial eigenstructure assignment -- which is concerned with assigning a subset of eigenvalues and/or eigenvectors while holding fixed the remainder of the spectrum -- also bears on the surgical eigenstructure assignment problem \cite{b7,b8,b9}. Indeed, Lu et al have implicitly provided an algorithmic solution for the surgical assignment problem using a partial eigenstructure assignment technique together with a standard pole-placement algorithm \cite{b7}, which further has a good computational performance; the result is however incomplete, in that it does not encompass the repeated-eigenvalue case. Several other efforts on partial eigenstructure assignment also provide reductions in computation, however they are focused on particular system structures (e.g. symmetric state matrices) or do not allow for verification for assignability \cite{b8,b9}.

%Of note, some recent works have considered eigenstructure assignment through state feedback, using approaches from geometric control theory \cite{g1,g2}. Specifically \cite{g2} has given significant insight into the assignable eigenstructure, the results developed thus far are also limited to distinct or non-defective eigenvalues cases.  

Our recent studies on designing secure control schemes for dynamics networks have involved surgical eigenstructure assignment, in particular requiring characterizations of assignability \cite{jointeig_app2,jointeig_app3}.  Based on this motivation, in this short article we pursue a complete treatment of the surgical assignment problem.  The main contributions of the article relative to the literature \cite{b2,b3,b7} are the following:

\begin{itemize}
\item A general condition for surgical assignment is obtained, which encompasses the non-distinct and defective eigenvalue cases.

\item A two-stage design method is outlined which generalizes the technique in \cite{b7} to the non-distinct and defective eigenvalue cases, while achieving the same level of computational complexity.

\item A direct, constructive approach is taken to develop the condition for assignability, which has the benefit of highlighting design flexibilities (non-uniqueness in solutions) at each stage.

\item The results are extended to address a more general {\em relaxed surgical assignment} problem, where overlapping subsets of both eigenvalues and eigenvectors are specified while other eigenvalues/eigenvectors remain unspecified.
\end{itemize}

We note that the results described here also complement and extend our recent work \cite{as1}, which pursued almost-surgical assignment (assignment to make eigenvalues arbitrarily close rather than exactly equal to specified locations in the complex plane).

\section{Main Result on Surgical Assignment}

In this note, the surgical assignment problem is solved by adapting the classical methods for full eigenstructure assignment developed by Moore et al \cite{b2,b3}, and using them in tandem with a generalization of the partial eigenstructure assignment approach given in \cite{b7}. The solution is complete, in the sense that only a condition on the specified subset of eigenvectors and generalized eigenvectors is required, while the remaining eigenvalues can be assigned without any limitation.

Formally, the following system is considered:
\begin{eqnarray} \label{open_loop_sys}
\mathbf{\dot{x}}= \mathbf{A}\mathbf{x}+\mathbf{B}\mathbf{u}
\end{eqnarray}
\noindent where $\mathbf{x} \in \mathbb{R}^n$ and $\mathbf{u} \in \mathbb{R}^m$ are the state and input.  The pair $(\mathbf{A},\mathbf{B})$ is assumed controllable. A static state feedback controller $\mathbf{u}=\mathbf{F} \mathbf{x}$ is applied. Our goal is to determine whether and how the feedback gain $\mathbf{F}$ can be designed, so that: 1)  $r$ closed-loop eigenvalues are placed at $\mathcal{L}_1= \{\lambda_1, \lambda_2, \cdots, \lambda_r\}$ and the corresponding eigenvectors or generalized eigenvectors are placed at $\mathcal{V}_1= \{ \mathbf{v}_1, \mathbf{v}_2, \cdots, \mathbf{v}_r \}$; and 2) the remaining $n-r$ eigenvalues are placed at $\mathcal{L}_2= \{\lambda_{r+1}, \lambda_{r+2}, \cdots, \lambda_n\}$. It is assumed that $\mathcal{L}_1$ and $\mathcal{L}_2$ are  self-conjugate sets. 

The feedback control gain is constructed as a sum of two terms ($\mathbf{F}= \mathbf{F}_0 +\mathbf{F}_1$), which are designed sequentially.  Specifically, the full eigenstructure assignment algorithm of Moore \cite{b2} and its generalizations \cite{b3} are adapted to assign $r$ eigenvalues and their associated eigenvectors/generalized eigenvectors to $\mathcal{L}_1$ and $\mathcal{V}_1$ respectively, using the gain $\mathbf{F}_0$ (provided that the target eigenvectors/generalized eigenvectors in $\mathcal{V}_1$ fall in certain subspaces). Then a generalization of the partial eigenstructure assignment method in \cite{b7} is developed encompassing generalized eigenvectors, which holds fixed the already assigned eigenstructure while allowing assignment of the remaining $n-r$ eigenvalues at will using the gain $\mathbf{F}_1$.

In \cite{b2}, Moore identified necessary and sufficient conditions on the target eigenvector set for full eigenstructure assignment with distinct eigenvalues; the case with repeated eigenvalues and generalized eigenvectors was subsequently addressed in \cite{b3}. Here, we assume that the partial set $\mathcal{V}_1$ satisfies commensurate conditions. Specifically,  it is assumed that (i) $\mathcal{V}_1$ is a set of $r$ linearly independent vectors within $\mathbb{C}^n$, (ii) the target eigenvectors and generalized eigenvectors in $\mathcal{V}_1$ associated with conjugate eigenvalues form conjugate vector spaces ($\mathbf{v}_k=\bar{\mathbf{v}}_i$ and $\lambda_k=\bar{\lambda}_i$), and (iii) for all $i= 1,2, \dots, r$ there are certain $\mathbf{z}_i \in \mathbb{C}^{q}$ for which (a) $[ (\mathbf{A}-\lambda_i \mathbf{I}_n) ~~ \mathbf{B} ] \begin{bsmallmatrix} \mathbf{v}_i \\
\mathbf{z}_i \end{bsmallmatrix} = \mathbf{0}$ when $\mathbf{v}_i$ is an eigenvector, or (b) when $\mathbf{v}_i$ is a generalized eigenvector then $[ (\mathbf{A}-\lambda_i \mathbf{I}_n) ~~ \mathbf{B} ] \begin{bsmallmatrix} \mathbf{v}_i \\
\mathbf{z}_i \end{bsmallmatrix} = \mathbf{v}_k; ~ k \in \{1,2, \dots, r\}$ where $\mathbf{v}_k$ is the previous eigenvector/generalized eigenvector  in the Jordan chain associated with a common eigenvalue in $\mathcal{L}_1$. We note that assumption (iii-b) implicitly requires that the eigenvector and generalized eigenvectors forming a Jordan chain are (partially) specified in sequence in $\mathcal{V}_1$. 

In analogy with \cite{b2,b3}, provided that assumption (iii) above holds, the closed-loop matrix $\mathbf{A+BF}_0$ is immediately seen to have eigenvalues $\mathcal{L}_1$ and corresponding eigenvectors and generalized eigenvectors $\mathcal{V}_1$ if the following equation holds:   
\begin{eqnarray} \label{part4_eqn1}
\mathbf{F}_0 ~ \mathbf{V}_{1} = \mathbf{Z}_{1}
\end{eqnarray}
where $\mathbf{V}_{1}=[\mathbf{v}_1 ~\mathbf{v}_2 \cdots ~ \mathbf{v}_r]$ and $\mathbf{Z}_{1} = [\mathbf{z}_1 ~\mathbf{z}_2 \cdots ~ \mathbf{z}_r]$. From assumption (i), the system of linear equations (\ref{part4_eqn1}) necessarily has a solution for $\mathbf{F}_0$. Indeed per assumption (ii), from standard manipulation of linear equations with conjugated coefficients, it can be shown that (\ref{part4_eqn1}) has a real solution $\mathbf{F}_0$. For $r<n$, the system of equations is underdetermined and has an infinite number of solutions; we choose $\mathbf{F}_0$ as any real solution of (\ref{part4_eqn1}). For this choice, $\mathbf{A+BF}_0$ has $r$ eigenvalues at $\mathcal{L}_1$ and corresponding eigenvectors and generalized eigenvectors at the desired $\mathcal{V}_1$, however the remaining eigenvalues and corresponding eigenspaces have not been specified.

It remains to design $\mathbf{F}_1$ to place the remaining eigenvalues of the closed-loop system \big($\mathbf{A+BF}=\mathbf{A+B}(\mathbf{F}_0+\mathbf{F}_1)$\big) at the target values in $\mathcal{L}_2$. This can be done via a generalization of the partial eigenstructure assignment method in \cite{b7}, which allows for non-distinct and defective eigenvalues. This generalization can also be achieved by applying a Jordan transformation to $\mathbf{A+B}\mathbf{F}_0$, but for computational reasons, following \cite{b7} we instead use an orthogonal transformation. To find the transformation, first consider the matrix $\mathbf{Q} \in \mathbb{R}^{n \times r}$ whose columns form an orthonormal basis for the subspace spanned by $\mathcal{V}_1$. (Such a $\mathbf{Q}$ exists since $\mathcal{V}_1$ is a self conjugate set of linearly independent vectors.) Then consider the matrix $\mathbf{W} \in \mathbb{R}^{n \times (n-r)}$ whose columns are an orthonormal basis for the orthogonal complement of the subspace spanned by $\mathcal{V}_1$. In this case, $\mathbf{T}=[\mathbf{Q}~~\mathbf{W}$] is an orthogonal matrix. 

The transformation specified by $\mathbf{T}$ will be used to design the gain $\mathbf{F}_1$. Prior to applying the transformation, it is helpful to develop an equivalence related to $\mathbf{Q}$, which is needed to characterize the spectrum upon transformation. To develop the equivalence, first note that $\mathbf{V}_{1}=[\mathbf{v}_1 ~\mathbf{v}_2 \cdots ~ \mathbf{v}_r]$ can be written as $\mathbf{V}_{1}= \mathbf{Q} \mathbf{R}$ where $\mathbf{R}$ is an $r \times r$ invertible matrix, since the columns of $\mathbf{Q}$ form an orthonormal basis for the subspace spanned by the columns of $\mathbf{V}_{1}$. Now, from the design of $\mathbf{F}_0$, we know that for all $i= 1,2, \dots, r$ either (a) $(\mathbf{A+BF}_0)\mathbf{v}_i= \lambda_i \mathbf{v}_i$ or (b) $(\mathbf{A+BF}_0)\mathbf{v}_i= \lambda_i \mathbf{v}_i+\mathbf{v}_k; ~ k \in \{1,2, \dots, r\}$. Given this, and given further that the Jordan chains in $\mathcal{V}_1$ are maintained (condition (iii)), the following equivalence holds:  $(\mathbf{A}+\mathbf{BF}_0)\mathbf{V}_{1}= \mathbf{V}_{1} \mathbf{\Lambda}_1$ where $\mathbf{\Lambda}_1$ is an $r \times r$ matrix whose eigenvalues are the same as those given in $\mathcal{L}_1$. Now note that $(\mathbf{A}+\mathbf{BF}_0) \mathbf{Q}= (\mathbf{A}+\mathbf{BF}_0) \mathbf{V}_{1} \mathbf{R}^{-1}= \mathbf{V}_{1} \mathbf{\Lambda}_1 \mathbf{R}^{-1}=$ $\mathbf{Q} \mathbf{R} \mathbf{\Lambda}_1 \mathbf{R}^{-1}$. From this, it follows immediately $\mathbf{A}_1=\mathbf{R} \mathbf{\Lambda}_1 \mathbf{R}^{-1}$ is a real matrix with eigenvalues equal to those given in $\mathcal{L}_1$. This observation along with the relation $(\mathbf{A}+\mathbf{BF}_0) \mathbf{Q}= \mathbf{Q} \mathbf{A}_1$ will play important role in our analysis.

Now to achieve the design, we select the gain $\mathbf{F}_1$ to have the form $\mathbf{F}_1=\mathbf{K}\mathbf{W}^T$, where $\mathbf{K} \in \mathbb{R}^{q \times (n-r)}$, and consider the spectrum of $\mathbf{A+BF}=(\mathbf{A}+\mathbf{B}\mathbf{F}_0)+\mathbf{B} \mathbf{F}_1$.  From the form of $\mathbf{F}_1$, it follows that  
$\mathbf{F}_1{\bf v}_i=0$ for $i=1, 2, \hdots, r$.  Hence, $\mathbf{A+BF}$ is verified to have eigenvalues at the target values specified in $\mathcal{L}_1$, with corresponding  eigenvectors and generalized eigenvectors as specified in  $\mathcal{V}_1$. To determine the remaining eigenvalues of $\mathbf{A+BF}$, notice that the matrix is similar to $\mathbf{T}^{-1} (\mathbf{A+BF}) \mathbf{T} = \mathbf{T}^T (\mathbf{A+BF}) \mathbf{T} $, and hence the matrices have identical eigenvalues. Next, using the fact that $\mathbf{Q}^T(\mathbf{A+BF}_0)\mathbf{Q} =$ $\mathbf{Q}^T \mathbf{Q} \mathbf{A}_1= \mathbf{A}_1$ (since $\mathbf{Q}^T \mathbf{Q} = \mathbf{I}_r$) and $\mathbf{W}^T(\mathbf{A+BF}_0)\mathbf{Q} =$ $\mathbf{W}^T \mathbf{Q} \mathbf{A}_1 = \mathbf{0}$  (since $\mathbf{W}^T \mathbf{Q} = \mathbf{0}$) we get
\begin{eqnarray} \label{trans_open_loop_sys}
\mathbf{T}^{-1} (\mathbf{A+BF}) \mathbf{T} = \begin{bmatrix} \mathbf{A}_1 & \mathbf{Q}^T(\mathbf{A+BF}_0)\mathbf{W} \\ \mathbf{0} & \mathbf{W}^T(\mathbf{A+BF}_0)\mathbf{W} \end{bmatrix} \nonumber \\
+\begin{bmatrix} \mathbf{Q}^T\mathbf{B}\\  \mathbf{W}^T\mathbf{B}\end{bmatrix}\begin{bmatrix} \mathbf{0} & 
\mathbf{K} \end{bmatrix}.
\end{eqnarray}
From this form, the remaining eigenvalues of $\mathbf{A+BF}$ are seen to be the eigenvalues of $\mathbf{A}_2+\mathbf{B}_2 \mathbf{K}$, where $\mathbf{A}_2= \mathbf{W}^T(\mathbf{A+BF}_0)\mathbf{W}$ and ${\mathbf{B}_2}=\mathbf{W}^T\mathbf{B}$. The eigenvalues $\mathbf{A}_2+\mathbf{B}_2 \mathbf{K}$ can be assigned at will by designing $\mathbf{K}$ if the pair $(\mathbf{A}_2,\mathbf{B}_2)$ is controllable. From the expression (\ref{trans_open_loop_sys}), however, it is also apparent that the pair $(\mathbf{A}_2,\mathbf{B}_2)$ is controllable
since $(\mathbf{A},\mathbf{B})$ has 
been assumed to be controllable.  Thus, we
have verified that the remaining eigenvalues of $\mathbf{A+BF}$ can be placed at will through solution of a lower-dimensional eigenvalue assignment problem, including at the locations specified in $\mathcal{L}_2$.  Thus, a complete solution to the surgical eigenstructure assignment problem has been developed via design of feedback gain $\mathbf{F}= \mathbf{F}_0 +\mathbf{K}\mathbf{W}^T$. We formalize this result as a theorem. 

\noindent \textbf{Theorem 1:}  \textit{Consider the system (\ref{open_loop_sys}), and assume that the system is controllable. A state feedback controller can be designed to: 1) place any $r$ closed-loop eigenvalues at any $\mathcal{L}_1$ $=\{\lambda_1, \lambda_2, \cdots, \lambda_r\}$ and their corresponding closed-loop eigenvectors/generalized eigenvectors at any $\mathcal{V}_1$ $= \{ \mathbf{v}_1, \mathbf{v}_2, \cdots, \mathbf{v}_r \}$, and 2) place remaining $n-r$ closed-loop eigenvalues at any $\mathcal{L}_2$ $=\{\lambda_{r+1}, \lambda_{r+2}, \cdots, \lambda_n\}$ provided that the following conditions hold: (i) $\mathcal{V}_1$ is a set of linearly independent vectors in $\mathbb{C}^n$; (ii)  $\mathcal{V}_1$ is a self conjugate set such that $\mathbf{v}_k= \bar{\mathbf{v}}_i$ and $\lambda_k = \bar{\lambda}_i$, where $i,k \in \{1,2, \cdots, r\}$; (iii) for all $i= 1,2, \cdots, r$ there are $\mathbf{z}_i \in \mathbb{C}^{q}$ for which (a) $[ (\mathbf{A}-\lambda_i \mathbf{I}_n) ~~ \mathbf{B} ] \begin{bsmallmatrix} \mathbf{v}_i \\
\mathbf{z}_i \end{bsmallmatrix} = \mathbf{0}$ when $\mathbf{v}_i$ is a target eigenvector, or (b) when $\mathbf{v}_i$ is a target generalized eigenvector then $[ (\mathbf{A}-\lambda_i \mathbf{I}_n) ~~ \mathbf{B} ] \begin{bsmallmatrix} \mathbf{v}_i \\
\mathbf{z}_i \end{bsmallmatrix} = \mathbf{v}_k; ~ k \in \{1,2, \cdots, r\}$ where $\mathbf{v}_k$ is the previous target eigenvector/generalized eigenvector to $\mathbf{v}_i$ in the Jordan chain associated with a common eigenvalue (i.e. $\lambda_k=\lambda_i$); and (iv) $\mathcal{L}_1$ and $\mathcal{L}_2$ are  self-conjugate sets.} \hfill  $\blacksquare$

Several remarks about the result are worthwhile:

\noindent \textbf{Remark 1:} Theorem 1 clarifies that surgical eigenstructure assignment is possible whenever the subset of specified eigenvectors/generalized eigenvectors fall in the corresponding vector spaces described by Moore and co-workers \cite{b2,b3}.  No further conditions on these or other eigenvectors/generalized eigenvectors are needed, and hence the condition for full eigenstructure assignment in \cite{b2,b3} is relaxed to an identical condition on only the specified eigenvectors/generalized eigenvectors.  We note that this relaxation does not follow immediately from the classical work on full eigenstructure assignment \cite{b2,b3}, since there is no guarantee that the remaining complement of eigenvectors/generalized eigenvectors can be chosen so that full set satisfies the additional required conditions (specifically, satisfaction of the additional vector-space constraints together with linear independence).  Thus, our analysis also indirectly gives insight into the classical work on full eigenstructure assignment \cite{b2,b3}. In particular, our result shows that, no matter how the independent eigenvectors/generalized eigenvectors associated with a subset of eigenvalues are chosen within the designated spaces (condition (iii) above), it is always possible to select eigenvectors/generalized eigenvectors associated with the remaining eigenvalues from the designated vector spaces to form a full complement. We formalize this in the following corollary which follows directly from the above theorem and Moore's result \cite{b3}.

\noindent \textbf{Corollary 1:}  \textit{Consider a set of $r$ complex numbers $\mathcal{L}_1$, a set of $r$ vectors $\mathcal{V}_1$ and a set of $(n-r)$ complex numbers $\mathcal{L}_2$. Assume  $\mathcal{L}_1$, $\mathcal{V}_1$ and $\mathcal{L}_2$ satisfy the conditions for surgical eigenstructure assignment given in Theorem 1, for a particular system  (\ref{open_loop_sys}). Then a set of $(n-r)$ vectors $\mathcal{V}_2$ can always be found such that the complex number set $\mathcal{L}=\mathcal{L}_1 \cup \mathcal{L}_2$ and the vector set $\mathcal{V}=\mathcal{V}_1 \cup \mathcal{V}_2$ satisfy the conditions given by Moore for full eigenstructure assignment via state feedback \cite{b2,b3}, for this system (\ref{open_loop_sys}).} \hfill  $\blacksquare$ 

The above corollary suggests an alternative way to achieve surgical assignment: rather than undertaking a two-stage design, one can directly compute eigenvectors/generalized eigenvectors associated with the eigenvalues in $\mathcal{L}_2$ in succession via linear-algebraic means to maintain the above conditions, with the guarantee that such eigenvectors/generalized eigenvectors can always be found.  Then the design algorithms in \cite{b2,b3} can be used directly. Thus, our analysis provides a guideline for guaranteed successive selection of unspecified eigenvectors/generalized eigenvectors, so that Moore's algorithm can be applied in solving the surgical eigenstructure assignment problem. The corollary generalizes a previous result on the assignability of sets of eigenvectors developed in the context of partial eigenstructure assignment \cite{b7}, to encompass the generalized-eigenvector case. The result here also differs from the previous studies, in that it falls out of a constructive design rather than coming from a linear-algebraic analysis.

\noindent \textbf{Remark 2:} The insight into assignable eigenspaces in Corollary 1 also falls out of the geometric characterizations of eigenstructure assignment developed in recent work \cite{g2}, however these characterizations are limited to the case with distinct eigenvalues.

%The corollary generalizes previous results on the assignability of sets of eigenvectors developed in the context of both partial eigenstructure assignment \cite{b7} and the geometric analysis of full eigenstructure assignment \cite{???}, which addressed the case of distinct eigenvalues.

\noindent \textbf{Remark 3:} The two-stage design presented here generalizes the design method given in \cite{b7}, to encompass the repeated- and defective- eigenvalue cases. It also allows flexibility in the controller design. Specifically, from our analysis, the gain of the controller achieving surgical assignment is given by $\mathbf{F}= \mathbf{F}_0 +\mathbf{K}\mathbf{W}^T$. Any real solution $\mathbf{F}_0$ of (\ref{part4_eqn1}) can be used; one choice is the least square solution $\mathbf{F}_0 =\mathbf{Z}_1\mathbf{V}_1^{\dagger}= \mathbf{Z}_1\mathbf{R}^{-1} \mathbf{Q}^{T}$. (In this expression, for computational reasons, it is preferred to replace each complex conjugate pair of column vectors in $\mathbf{V}_1$ and $\mathbf{Z}_1$ by their corresponding real and imaginary parts.) On the other hand, $\mathbf{K}$ can be obtained by applying any convenient eigenvalue placement technique (e.g. \cite{pole_1,pole_2}) to place the eigenvalue of the system $(\mathbf{A}_2,\mathbf{B}_2)$ at $\mathcal{L}_2$. This two-stage design can reduce the dimension of matrix computations relative to Moore's algorithm, in the same way as \cite{b7}. In particular, the computational complexity of the two-stage design can be reduced to $O(nr^2)+O(r^3)+O(rn^2)+O(n(n-r)^2)+O((n-r)n^2)+ O((n-r)^4)$ whereas the computational complexity of adapted Moore's algorithm for surgical assignment (as outlined in Remark 1) is $O((n-r)n^3)$ \cite{b7}.

%For instance,  accounting for matrix inversions only as the dominant computational cost, we see that the comutational complexity of the two stage design scales with $(r^3)+(n-r)^3$, whereas the computation complexity of Moore's algorithm scales with $n^3$.}

\noindent \textbf{Remark 4:} The conditions given in the Theorem 1 are also necessary for surgical eigenstructure assignment when $\mathcal{L}_1$ and $\mathcal{L}_2$ are  self conjugate sets and $\mathcal{L}_1 \cap \mathcal{L}_2 =\emptyset$. This follows directly from \cite{b2,b3} as the conditions given by Moore are also necessary. 

\noindent \textbf{Remark 5:} The analysis can be generalized to encompass the uncontrollable case, with the uncontrollable eigenvalues and associated left eigenvectors simply remaining fixed.  Details are omitted.

\section{Generalized Surgical Assignment}

A more general version of the surgical eigenstructure assignment problem is considered, wherein possibly-overlapping subsets of eigenvectors and eigenvalues are assigned, while the remaining ones are left free. We refer to this generalization as the {\em relaxed surgical-eigenstructure assignment} problem. Specifically, we consider an eigenstructure assignment problem where a controller must be designed to place: 1) $r_1$ closed-loop eigenvalues at $\mathcal{L}_{g1}= \{\lambda_1, \lambda_2, \cdots, \lambda_{r_1}\}$ and the corresponding closed-loop eigenvectors or generalized eigenvectors at $\mathcal{V}_{g1}= \{\mathbf{v}_1, \mathbf{v}_2, \cdots, \mathbf{v}_{r_1} \}$; 2) $r_2$ closed-loop eigenvalues at $\mathcal{L}_{g2}= \{\lambda_{r_1+1}, \lambda_{r+2}, \cdots, \lambda_{r_1+r_2}\}$; 3) $r_3$ closed-loop eigenvectors at $\mathcal{V}_{g3}= \{\mathbf{v}_{r_1+r_2+1}, \mathbf{v}_{r_1+r_2+2}, \cdots, \mathbf{v}_{r_1+r_2+r_3}\}$ where $r_1+r_2+r_3 \leq n$.  This kind of eigenstructure assignment problem arises in settings where certain eigenvectors require assignment (e.g. to prevent estimation of state information by an adversary, or to shape participation in certain modal dynamics), but there is flexibility in the corresponding eigenvalue locations.

Here, conditions on the eigenstructure are developed, such that relaxed surgical assignment can be achieved  through state feedback.  A design algorithm is also presented. The conditions and algorithm derive from the surgical eigenstructure assignment results presented above. For the derivation, first we make some further observations about the assignable eigenvectors associated with each eigenvalue. Specifically, we note that Moore's third condition characterizing the space of assignable eigenvectors can be rewritten as follows: an eigenvector can be assigned in the direction $\mathbf{v}_i$ if the following equation has a solution for $\mathbf{z}_i$ and $\lambda_i$:
\begin{equation} \label{eigenvector_assign}
[\mathbf{B} ~~ \mathbf{v}_i] 
\left[\begin{array}{c}
-\mathbf{z}_i \\
\lambda_i \end{array}\right]  = \mathbf{A}\mathbf{v}_i.
\end{equation}
It is clear that the equation (\ref{eigenvector_assign}) has a solution $\begin{bsmallmatrix} -\mathbf{z}_i \\ \lambda_i \end{bsmallmatrix}$ if and only if $\mathbf{A}\mathbf{v}_i$ is in the range space of $[\mathbf{B} ~~ \mathbf{v}_i]$. Therefore, $\mathbf{v}_i$ can be assigned as a closed-loop eigenvector if and only if $\mathbf{A}\mathbf{v}_i$ is in the range space of $[\mathbf{B} ~~ \mathbf{v}_i]$. From the linear form of Equation (\ref{eigenvector_assign}), it further follows that the closed-loop eigenvalue $\lambda_i$ corresponding to an assignable closed-loop eigenvector $\mathbf{v}_i$ is either unique or arbitrary in $\mathbb{C}$. Specifically, $\lambda_i$ can be chosen arbitrarily if and only if both $\mathbf{v}_i$ and $\mathbf{Av}_i$ are in the range space of $\mathbf{B}$. Also, if $\begin{bsmallmatrix} -\mathbf{z}_i \\ \lambda_i \end{bsmallmatrix}$ satisfies (\ref{eigenvector_assign}) for $\mathbf{v}_i$, then $\begin{bsmallmatrix} -\bar{\mathbf{z}}_i \\ \bar{\lambda}_i \end{bsmallmatrix}$ satisfies (\ref{eigenvector_assign}) for $\bar{\mathbf{v}}_i$.

From the above discussion and Theorem 1, the following result on relaxed surgical-eigenstructure assignment follows immediately.

\noindent \textbf{Theorem 2:} \textit{Consider the system (\ref{open_loop_sys}), and assume that the system is controllable. Consider design of a state feedback controller to solve the relaxed surgical-eigenstructure assignment problem, i.e. to place: 1) $r_1$ closed-loop eigenvalues at $\mathcal{L}_{g1}= \{\lambda_1, \lambda_2, \cdots, \lambda_{r_1}\}$ and the corresponding closed-loop eigenvectors or generalized eigenvectors at $\mathcal{V}_{g1}= \{\mathbf{v}_1, \mathbf{v}_2, \cdots, \mathbf{v}_{r_1} \}$; 2) $r_2$ closed-loop eigenvalues at $\mathcal{L}_{g2}= \{\lambda_{r_1+1}, \lambda_{r+2}, \cdots, \lambda_{r_1+r_2}\}$; 3) $r_3$ closed-loop eigenvectors at $\mathcal{V}_{g3}= \{\mathbf{v}_{r_1+r_2+1}, \mathbf{v}_{r_1+r_2+2}, \cdots, \mathbf{v}_{r_1+r_2+r_3}\}$ where $r_1+r_2+r_3 \leq n$.  A state feedback controller can be designed for this purpose provided that the following conditions hold: (i) $\mathcal{V}_{g1} \cup \mathcal{V}_{g3}$ is a set of linearly independent vectors in $\mathbb{C}^n$; (ii)  $\mathcal{V}_{g1}$ is a self conjugate set such that $\mathbf{v}_k= \bar{\mathbf{v}}_i$ and $\lambda_k = \bar{\lambda}_i$ where $i,k \in \{1,2, \cdots, r_1\}$; (iii) $\mathcal{V}_{g3}$ is a self conjugate set; (iv) for all $i= 1,2, \cdots, r_1$ there are certain $\mathbf{z}_i \in \mathbb{C}^{q}$ for which (a) $[ (\mathbf{A}-\lambda_i \mathbf{I}_n) ~~ \mathbf{B} ] \begin{bsmallmatrix} \mathbf{v}_i \\
\mathbf{z}_i \end{bsmallmatrix} = \mathbf{0}$ when $\mathbf{v}_i$ is a target eigenvector or, (b) when $\mathbf{v}_i$ is a target generalized eigenvector then $[ (\mathbf{A}-\lambda_i \mathbf{I}_n) ~~ \mathbf{B} ] \begin{bsmallmatrix} \mathbf{v}_i \\
\mathbf{z}_i \end{bsmallmatrix} = \mathbf{v}_k; ~ k \in \{1,2, \cdots, r_1\}$ where $\mathbf{v}_k$ is the previous target eigenvector/generalized eigenvector to $\mathbf{v}_i$ in the Jordan chain associated with a common eigenvalue (i.e. $\lambda_k=\lambda_i$); (v) $\mathbf{A}\mathbf{v}_i$ is in the range space of $[\mathbf{B} ~~ \mathbf{v}_i]$ for all $i= r_1+r_2+1, r_1+r_2+2, \cdots, r_1+r_2+r_3$; and (vi) $\mathcal{L}_{g1}$ and $\mathcal{L}_{g2}$ are self-conjugate sets.} \hfill  $\blacksquare$ 

When the conditions in Theorem 2 are satisfied, the feedback controller design can be done in the following manner. First, the self conjugate set $\mathcal{L}_{g3}$ is found for the vector set $\mathcal{V}_{g3}$ via solving (\ref{eigenvector_assign}), which yields closed-loop eigenvalues associated with the target closed-loop eigenvectors specified in $\mathcal{V}_{g3}$. For $r_1+r_2+r_3 < n$ any self conjugate set of $n-(r_1+r_2+r_3)$ complex numbers $\mathcal{L}_{g4}$ is then chosen for the remaining closed-loop eigenvalues which have not been specified yet. Then by considering $\mathcal{L}_1=\mathcal{L}_{g1} \cup \mathcal{L}_{g3}$, $\mathcal{V}_1=\mathcal{V}_{g1} \cup \mathcal{V}_{g3}$ and $\mathcal{L}_2=\mathcal{L}_{g2} \cup \mathcal{L}_{g4}$, and applying the surgical eigenstructure design methods presented previously, the gain $\mathbf{F}$ is found that achieves the relaxed surgical-eigenstructure assignment.  We notice that this algorithmic construction also serves to prove Theorem 2.

\section{Example}

We present a simple numerical example to illustrate the application of our results.

\underline {Example}: Consider a linear system with state matrix $\mathbf{A}=[1, 2, 1, 2;$ $1, 1, 0, 1;$ $1, 0, 2, 0; 1, 0, 0, 1]$ and input matrix $\mathbf{B}= [1, 0; 0, 1;$ $0,  0; 0,  0]$. The designer's goal is to assign the closed loop eigenvalues at $\{0,1,3,5\}$ via a state feedback controller. Additionally the designer seeks to place the closed loop eigenvectors corresponding to the first three eigenvalues at $[2; 0; -1; -2],$ $[0;  1;  0;  0]$ and $[2;  0;  2;  1]$ respectively. Now considering $\mathcal{L}_1=\{0,1,3\}$, $\mathcal{V}_1=\{[2; 0; -1; -2],$ $[0;  1;  0;  0],$ $[2;  0;  2;  1]\}$ and $\mathcal{L}_2= \{5\}$ we notice that $\mathcal{L}_1$, $\mathcal{V}_1$ and $\mathcal{L}_2$ satisfy all the conditions of Theorem 1. Therefore, a gain matrix $\mathbf{F}$ exists that yields the target surgical eigenstructure assignment. For instance, using our design method we find $\mathbf{F}=[4, -2, -7, 6;  2.5817, 0, -7.1634, 6.1634]$ for which the closed loop system has the target eigenstructure. From Theorem 1, it is clear that surgical assignment is  possible when the target location of the fourth closed-loop eigenvalue is any real value (including $0, 1$ and $3$).

Now consider a different case where the designer wants the same surgical eigenstructure  assignment except the third eigenvector is placed at $[1;  1;  0;  0]$ instead of $[2;  0;  2;  1]$. Here we again consider $\mathcal{L}_1=\{0,1,3\}$, $\mathcal{L}_2= \{5\}$ and $\mathcal{V}_1=\{[2; 0; -1; -2],$ $[0;  1;  0;  0],$ $[0;  1;  1;  0]\}$. However, for this case the third eigenvalue and eigenvector pair does not satisfy the third condition of Theorem 1. Hence, surgical assignment cannot be achieved in this case. However, if the designer relaxes the location of third eigenvalue, then this relaxed surgical-eigenstructure assignment will be possible since $\mathcal{L}_1=\{0,1\}$, $\mathcal{V}_1=\{[2; 0; -1; -2]$ $[0;  1;  0;  0] \}$, $\mathcal{L}_2= \{5\}$,  and $\mathcal{V}_3=\{[0;  1;  1;  0]\}$ satisfy all the conditions of Theorem 2. Using our design method we find that the gain $\mathbf{F}=[3, -2, -1, 2; 1.2211, 0, 1, 0.7211]$ yields this relaxed surgical-eigenstructure assignment. For this gain matrix, the third closed loop eigenvalue is placed at $2$ instead of $3$ so that its corresponding eigenvector is the target vector $[0;  1;  1;  0]$. However, if the designer wanted the third eigenvector to be placed at $[1;  1;  1;  1]$, in that case the relaxed surgical-eigenstructure assignment would not be possible since this vector does not satisfy assignability condition for any eigenvalue.

\end{document}